\documentclass[manuscript,nonacm]{acmart}

\usepackage{graphicx}
\usepackage[normalem]{ulem}
\useunder{\uline}{\ul}{}
\usepackage{multirow}
\usepackage{adjustbox}
\usepackage{caption,subcaption,graphicx}

\newcommand{\red}[1]{\textcolor{black}{#1}}
\colorlet{RED}{red}

\AtBeginDocument{%
  \providecommand\BibTeX{{%
    \normalfont B\kern-0.5em{\scshape i\kern-0.25em b}\kern-0.8em\TeX}}}

\setcopyright{acmcopyright}
\copyrightyear{2023}
\acmYear{2023}
\acmDOI{10.1145/1122445.1122456}

\acmConference[ACM TOCHI '24 Submission]{ACM TOCHI '24 Submission}{2024}{}
\acmBooktitle{ACM TOCHI '24Submission}
\acmPrice{15.00}
\acmISBN{978-1-4503-XXXX-X/18/06}

\begin{document}

\title{Investigating an Intelligent System to Monitor \& Explain Abnormal Activity Patterns of Older Adults}
%\title[Investigating Opportunities of Ambient-Assisted Living Technologies to Monitor Abnormal Daily Activities of Older Adults]{Investigating Opportunities of Ambient-Assisted Living Technologies to Monitor Abnormal Daily Activities of Older Adults: A Qualitative Study with Older Adults and Caregivers}

\author{Min Hun Lee}
\email{mhlee@smu.edu.sg}
\affiliation{%
  \institution{Singapore Management University}
  %\streetaddress{P.O. Box 1212}
  \city{Singapore}
  \country{Singapore}
}

\author{Daniel P. Siewiorek}
\affiliation{%
  \institution{Carnegie Mellon University}
   \city{Pittsburgh}
  \country{USA}}
\email{dps@cs.cmu.edu}

\author{Alexandre Bernardino}
\affiliation{%
  \institution{Instituto Superior Técnico}
  %\streetaddress{30 Shuangqing Rd}
  \city{Lisboa}
  \country{Portugal}}
\email{alex@isr.tecnico.ulisboa.pt}

\renewcommand{\shortauthors}{M.H. Lee, et al.}

\begin{abstract}
Despite the growing potential of older adult care technologies, the adoption of these technologies remains challenging. In this work, we conducted a focus-group session with family caregivers to scope designs of the older adult care technology. We then developed a high-fidelity prototype and conducted its qualitative study with professional caregivers and older adults to understand their perspectives on the system functionalities. This system monitors abnormal activity patterns of older adults using wireless motion sensors and machine learning models and supports interactive dialogue responses to explain abnormal activity patterns of older adults to caregivers and allow older adults proactively sharing their status with caregivers for an adequate intervention. Both older adults and professional caregivers appreciated that our system can provide a faster, personalized service while proactively controlling what information is to be shared through interactive dialogue responses. We further discuss other considerations to realize older adult technology in practice.
\end{abstract}



\maketitle

\section{Introduction}
The number of people aged 60 or over is expected to increase to 2.1 billion in 2050 and 3.1 billion in 2100 \cite{desa2017united}. In the face of an aging population and increasing chronic disease, and anticipated shortages of healthcare workers, our society requires new efficient ways to deliver healthcare services \cite{dexter2010preparing}. Researchers have explored older adult care technologies \cite{rashidi2012survey} {to support their independent living and improve their quality of life}. The idea of such technologies \red{\cite{rashidi2012survey,blackman2016ambient,cardinaux2011video} ranges from a personal alarm system worn by an older adult for raising the alarm in the case of an emergency situation, technologies that provide physical assistance \cite{rashidi2012survey,mukai2010development,erickson2018deep,gallenberger2019transfer} or companionship \cite{sidner2018creating}
, and intelligent systems that analyze sensor data to detect potential environmental hazards in the home environment and the abnormal activity patterns of an older adult \cite{rashidi2012survey,blackman2016ambient}. Among various technologies, this work focuses on an intelligent system that} analyzes sensor data to recognize activities of daily living \cite{ordonez2013activity,fortin2015exploiting,nweke2018deep}, detect a dangerous or abnormal situation \red{(e.g. being idle for a long time, leaving home and wandering at abnormal hours, etc.) that may be indicative of changes in health status} \cite{shin2011detection,ordonez2015sensor}, and alert an older adult's abnormal situation to a caregiver to provide an intervention if necessary \cite{rashidi2012survey,blackman2016ambient}. 

\red{An intelligent system in the home environment that monitors activities of daily living might contribute to an increased sense of safety for older adults' independent living \cite{yusif2016older}. Previous studies have shown} the value of reviewing observations of activities of daily living \cite{lee2015sensor} and providing contextual information \cite{wu2018understanding} to improve the self-awareness of older adults and the potential of a conversational agent to assist daily activities (e.g. alarms during cooking, getting weather information) \cite{zubatiy2021empowering}. However, the adoption of such technologies remains challenging due to several issues \cite{heart2013older,ehrenhard2014market,yusif2016older}: lack of usability and trust \cite{yusif2016older}, lack of options to control these technologies \cite{berridge2022control} and privacy concerns \cite{ehrenhard2014market}. \red{Even if researchers have explored} complex algorithms for more accurate \red{detections} of activities of daily living \cite{ordonez2016deep,nweke2018deep} and abnormal events \cite{ordonez2015sensor}, the systems with complex algorithms operate as black-box systems. \red{They} only provide \red{the end-users} a one-directional notification on their \red{detected abnormal daily living events, which can further exacerbate the trust, control, and overall user experience of the system.% limited interaction on their predictions. 
}

%To address this gap, we 
In this paper, we present our investigation on an intelligent system \red{that monitors activities of daily living} for older adult care to improve its user experience and control. First, we conducted a focus group session with family caregivers to specify the design spaces of a system. Based on the findings of the focus group session, we implemented an intelligent system that can monitor abnormal patterns of activities of daily living and provide interactive dialogue responses that explain an abnormal event to a caregiver and allow older adults to proactively control sharing their status for a personalized intervention. In addition, we conducted a qualitative study with older adults and professional caregivers after showing video demonstrations of the system to understand their perspectives on the system designs and functionalities. 

Our qualitative study showed that both professional caregivers and older adults perceived opportunities of the system to provide faster attentive, personalized care services while reducing cognitive burdens and workloads of caregivers. Caregivers and older adults appreciated the value of our system with interactive dialogues to balance the boundary management of information to be shared: caregivers can review the explanations of abnormal events and elicit additional information from an older adult while older adults can proactively control if and what information can be shared. In addition, they elaborated on several limitations of the system (e.g. questionable system performance, interaction approaches, and financial and policy aspects). Based on the findings, we discuss other considerations (e.g. explainability, interactivity, a new benchmark dataset, human-centered evaluations, multimodal sensing, omnipresent interactions, etc.) to make older care technologies more controllable and adoptable in practice. 

In summary, this work contributes to designs on how to improve the user experiences and control of older adult technologies for independent living of older adults. Our work advances the growing body of research in older adult care technologies that engage with stakeholders (e.g. caregivers and older adults) for human-centered designs and discuss the issue of how to improve the end users (e.g. older adults and caregivers)' control and experiences on these technologies.

\section{Related Work}

\subsection{Older Adult Care Technologies}\label{related-aal-tech}
\subsubsection{Implementations and Applications}
With the growth of the aging population \cite{nations2017department}, researchers have explored various older adult care technologies to support more independent living of older adults while improving the workflows of caregivers \cite{rashidi2012survey}. The implementation of these systems involves deploying sensors in an environment to collect continuous information on the occupants and their health (e.g. daily activities, heart rate, body temperature) \red{ \cite{rashidi2012survey,blackman2016ambient,cardinaux2011video}}. 
These technologies can be realized in a variety of applications. For instance, these systems can provide reminders on taking a medication \cite{lee2015sensor} or cognitive assistance on daily activities \cite{ha2014towards} or exercises \cite{lee2020towards}. In addition, these systems can detect the occurrence of abnormal or dangerous situations (e.g. a fall \cite{stone2014fall,wu2018understanding,AppleSupport2023}, night-time wandering \cite{vuong2011feasibility}) and alert a caregiver or health professional \cite{AltumView2023} to provide the immediate, necessary support. In this paper, we focus on the application to monitor the daily activities of older adults and detect abnormal events for their safe and independent living. 

To monitor the daily activities of older adults, researchers have explored various sensors, such as depth video sensors \cite{cardinaux2011video,jalal2014depth}, wall-mounted radio-frequency identifications (RFIDs) \cite{stikic2008adl,fortin2015exploiting}, wearable sensors \cite{rashidi2012survey,nweke2018deep}, and wireless infrared sensors \cite{shin2011detection,ordonez2013activity,ordonez2015sensor,forkan2015context}. Given sensor data, researchers explore various machine learning algorithms to recognize daily activities and detect abnormal events \cite{shin2011detection,ordonez2013activity,ordonez2015sensor,nweke2018deep}. 

\subsubsection{User Studies}
In addition, researchers have conducted studies to better understand the benefits and challenges of technologies for older adults and caregiving. \red{For instance,} Lee and Dey \cite{lee2015sensor} conducted studies with older adults and described that reviewing observations of daily living from ubiquitous sensors can help older adults to enhance their self-awareness of their abilities and assist clinicians to collect useful information to guide their care of the older adult. Wu and Munteanu \cite{wu2018understanding} conducted an evaluation study of wearable sensor-based fall risk assessment with older adults and described the importance of providing contextual information to help the user understand their situation better. Zubatiy et al. \cite{zubatiy2021empowering} collected interaction data of a conversational agent, Google Home Hub from older adults with mild cognitive impairment and care partners and described the potential of a conversational agent to empower daily life activities (e.g. alarms to help cooking and getting weather information).

There has been a growing body of research on older adult care technologies and commercialized services (e.g. fall detection feature of Apple Watch \cite{AppleSupport2023}). These technologies have great promises to revolutionize healthcare services. However, the adoption of these technologies is still very low due to several issues \cite{heart2013older,ehrenhard2014market,yusif2016older}. These issues include usability and trust of these systems \cite{yusif2016older,lee2015perspective,pal2018internet,hoque2017understanding} and concerns on lack of control agency by older adults \cite{berridge2022control} and privacy \cite{ehrenhard2014market,claes2015attitudes,yusif2016older}. As these systems often involve a complex algorithm \cite{ordonez2013activity,ronao2016human,ordonez2016deep,wang2019deep} to analyze sensor data, they perform as black-box systems that the user cannot understand why a system provides a certain output (e.g. detecting an abnormal event). Thus, both older adults and caregivers cannot have \red{trustworthy} usage of these systems. In addition, these systems might not perform well in the real-world \cite{vacher2011development,cardinaux2011video,rashidi2012survey}.

In contrast to prior work (Section \ref{related-aal-tech}) that focuses on improving the performance of a system to recognize the \red{activities} of daily living and abnormal events using complex algorithms \cite{ordonez2013activity,ronao2016human,ordonez2016deep,wang2019deep}, we engaged with older adults and family and professional caregivers to explore how an older adult care technology can be designed to improve the end-user's control.
%design and evaluate a human-centered, intelligent system for older adult care. 
Specifically, this system explores decision trees with contextual features of an activity (e.g. the duration, frequency, starting hour of activity, and transitions between activities) to explain an abnormal activity pattern of an older adult. Also, this system leverages interactive dialogue responses to support caregivers to elicit additional information from an older adult while allowing older adults to proactively control what information is to be shared.

\subsection{\red{Understanding Needs of} Older Adults and Caregivers}
One important factor in making a system more adoptable is to improve the user expectation of the potential value of a system \cite{heart2013older,lee2015perspective,yusif2016older,hoque2017understanding,pal2018internet}. Engagement with the users to understand their needs and preferences to design care technology is an active research topic. Human-computer interaction (HCI) research has made great strides in technology design and evaluation of care technologies through engagements with the users \cite{harrington2018designing,hong2016care,seo2021learning,berry2019supporting,guan2021taking,berridge2022control}. For instance, prior studies have explored technology designs for an effective communication between care providers and adults with chronic conditions \cite{berry2019supporting}, child patients \cite{hong2016care,seo2021learning}, or people with dementia \cite{guan2021taking}. Harrington et al. had design sessions with older adults and provided recommendations for mobile fitness technologies \cite{harrington2018designing}. Kuoppam{\"a}ki et al. \cite{kuoppamaki2021designing} analyzed six video recordings of older adults preparing a meal in their kitchen to provide design implications for assistive kitchen technologies for aging in place. In addition to these design studies on specific contexts \cite{harrington2018designing,hong2016care,seo2021learning,berry2019supporting,guan2021taking}, previous studies also discussed that even if older adults considered the control of technologies is a very important factor for their acceptance \cite{nurgalieva2019information,berridge2022control}, existing monitoring technologies have the limitation of not being customizable to accommodate personal preference \cite{vines2013making,lee2022enabling}.

\red{In addition,} even if prior design works engaged with older adults or/and caregivers to discuss design considerations for various applications (e.g. mobile fitness \cite{harrington2018designing} and assistive kitchen \cite{kuoppamaki2021designing}) and the importance of improving communication between care providers and adults with chronic conditions \cite{berry2019supporting} or dementia \cite{guan2021taking} and controlling older adult care technologies \cite{nurgalieva2019information,berridge2022control}, it remains unclear about detail design specifications how we can provide adequate control of older adult care technologies. To address this gap, we utilized interactive dialogue responses as a means of controlling older care technologies as previous studies reported that many older adults felt natural and empowered to interact with any form of smart assistants in their home \cite{zubatiy2021empowering}. We conducted a qualitative study of our system designs and functionalities with older adults and professional caregivers to seek the perceived opportunities and limitations for creating more controllable older adult care technologies. Our work brings insights into the potential of dialogue responses to facilitate communications between caregivers and older adults and support older adults' control over what information can be shared. In addition, we also discuss other considerations (e.g.
explainability, interactivity, a new benchmark dataset, human-centered evaluations, multimodal sensing, omnipresent
interactions, etc.) to make older care technologies more controllable and adoptable in practice.

\section{Study on a Human-Centered, Intelligent System for Older Adult Care}

In this work, we engaged with family and professional caregivers and older adults to design a human-centered intelligent system for older adult care through the following three research activities.
First, we conducted a focus-group session with four family caregivers to understand the practices and challenges of caregivers and specified the design spaces of the system. Based on the findings of the focus-group session, we implemented the high-level system prototype to illustrate its functionalities. In addition, we conducted a qualitative study of the system designs and functionalities with five professional caregivers and five older adults. The purpose of this qualitative study was not meant to be a comprehensive evaluation of our system, but to collect early feedback on our system designs and functionalities. During the qualitative study, we showed the video demonstration of the system to prompt discussions on the opportunities and limitations of the system and asked for new ideas for more empowering and controllable interactions with older adult care technologies.

\section{Focus group session with caregivers}
The objectives of an initial focus group with family caregivers \red{were} to learn about their experiences and challenges during the care services of an older adult and specify the design spaces of an intelligent system. For the focus group session, we recruited four family caregivers, who have had experiences to provide caregiving for their parents or grandparents (C1 - C4 in Table \ref{tab:participants-caregivers-focus}).  

\begin{table}[htp]
\centering
\caption{Demographics information of caregivers of the focus group session for the scope and design}
\label{tab:participants-caregivers-focus}
\resizebox{0.85\textwidth}{!}{%
\begin{tabular}{cccc} \toprule
\textbf{PID} & \textbf{Studies} &
  \textbf{Gender} &
  \textbf{Caregiving Experience} \\ \midrule
C1 & Scope \& Design &
  Male &
  \begin{tabular}[c]{@{}c@{}}A primary caregiver for both parents\\(one with mild cognitive impairment \& urination issues; one with Parkinson's disease\end{tabular} \\ \midrule
C2 & Scope \& Design &
  Female &
  \begin{tabular}[c]{@{}c@{}}A primary caregiver for both parents and a secondary caregiver for grandparents\\(one with anxiety disorder; one requires meal assistance; others: N/A)\end{tabular}  \\ \midrule
C3 & Scope \& Design &
  Male &
  \begin{tabular}[c]{@{}c@{}}A primary caregiver for both parents\\(one with stroke and swallowing issue; one: N/A)\end{tabular}  \\ \midrule
C4 & Scope \& Design &
  Female &
  \begin{tabular}[c]{@{}c@{}}A primary caregiver for both parents\\(one with cancer; one with falling \& alcoholism)\end{tabular} \\ \bottomrule
\end{tabular}%
}
\end{table}

A researcher then moderated a semi-structured focus group session with the following three questions to center conversations on learning practices of the care service for an older adult:

\begin{enumerate}
    \item What is the process of determining whether an older adult is okay or not?
    \item What signals that you use to make a decision for an older adult?
    \item How do you react to those signals and what's the process of interacting with an older adult?
\end{enumerate}

For the data analysis, we transcribed the audio recordings of the focus group session and followed an iterative coding process \cite{braun2006using}. First, we generated initial codes from the structured topics of the interviews. After reviewing transcripts, the researchers individually generated codes and findings and iteratively improved the codes of the transcripts with the research team. In the following subsections, we described the thematic analysis and findings of the focus-group session: practices and challenges of providing a care service for an older adult and high-level designs of an intelligent system for older adult care. 

\subsection{Practices and challenges of the care services}\label{sect:focusgroup-practices}
The care services of an older adult broadly start with checking the status of an older adult and providing an adequate intervention if necessary.

\subsubsection{Regular check-ins through visiting or phone calls}
%\hfill
%
%\noindent
For checking the status of an older adult, caregivers mainly rely on visiting or phone calls. \textit{``What we do is just regular check-ins by speaking them on the phone and getting a sense of how things work''} (C1). Also, caregivers need to visit an older adult periodically and spend time together to understand the activities of an older adult as \textit{``it is hard to get an older adult on the phone''} (C2). Such regular check-ins through visiting or phone calls \textit{``are for a very affection. Sometimes, both of us are relatively sort of bored and feel stressed. Although I would like to move away, but she felt trapped''} (C3).

%become \textit{``extremely tiring and feeling stressful. Although I would love to move away, but she felt trapped''} (C3).

\subsubsection{Speculating Activities to check the status of an older adult}
%\hfill
%
%\noindent
During visiting or phone calls, caregivers typically speculate on the daily activities of an older adult to determine the necessity of intervention. Especially, caregivers get a sense of when an older adult \textit{``wake up, eat foods, spend spare time, and leave the house''} (C1). Caregivers then check any deviations of activity that occur, such as \textit{``canceling the regular dinner''} (C2). Also, C3 described the importance of identifying the situation or signs of falling: \textit{``my father wasn't telling me anything about falling, but he told my boyfriend later and he wasn't able to get up for an hour (...) all of our emergency room visits were from his falling''} (C3). Although we observed several daily activities that are critical to monitor the status of an older adult, \textit{``we cannot enumerate the list as it is very, very individualized.''} (C3). C3 also described the importance to get a sense of \textit{``if the person is the same as the person you spoke with the last time''} (C3) through asking a question.
Once a caregiver observes any situation (e.g. \textit{```older adults cannot have a meal themselves''} - C2), caregivers will then figure out what services they could provide an older adult from social workers or a hospital. 

\subsubsection{Preserve the independence and dignity of an older adult}
%\hfill
%
%\noindent
All caregivers in the focus group session highlighted the importance of preserving the independence and dignity of an older adult. \textit{``You really think this isn't a pet that you are just trying to keep alive. This is a person''} (C1). Sometimes, when a caregiver provides a service \textit{``that is not necessary, my dad felt the invasion of his privacy''} (C4). In addition, \textit{``most problems that older adults would get embarrassing. My dad carrying around a urine bag for several months'' } (C3). Some older adults become \textit{``stressful about receiving a service from a different person each time''} (C4). Thus, it is critical to make a balanced decision to provide an adequate intervention that preserves the independence and dignity of an older adult.

\subsection{High-Level Designs of an Intelligent System}
Based on our findings from the focus group session, we identified high-level designs of an intelligent system to improve the care practices of an older adult (Table \ref{tab:findings-requirements}). First, the system should recognize the daily activities of an older adult and detect abnormal events to reduce the caregiver's check-in visits or phone calls. During the focus-group session, we identified daily activities that are commonly used to check the status of an older adult. These activities include \textit{``Waking up, Sleeping, Eating (Breakfast, Lunch, Dinner, Snack), SpareTime, Leaving, Toileting, and Idle/Falling''}. To preserve the independence and dignity of an older adult, the system should \textit{``reduce the installation of a camera everywhere''} (C1) and prioritize the usage of non-visual sensors.

In addition, as an older adult might have different life patterns and preferences on what information can be shared, the system should be able to provide a means of personalization and control. Specifically, we observed that a caregiver usually engages in a conversation and asks a specific, tailored question to an older adult to seek out further information on an older adult for analyzing his/her status and determining an adequate intervention while preserving his/her dignity. Thus, the system should provide a way to facilitate communication between an older adult and a caregiver. 

\begin{table}[htp]
\centering
\caption{List of findings from the focus-group and the corresponding high-level designs of an intelligent system for an older adult}
\label{tab:findings-requirements}
\resizebox{\textwidth}{!}{%
\begin{tabular}{ll} \\ \toprule
\multicolumn{1}{c}{\textbf{Findings from the focus-group}}  & \multicolumn{1}{c}{\textbf{High-Level Designs of a system}}                                     \\ \midrule
F1. Limitations of regular check-ins on the phone or visiting &
  \begin{tabular}[c]{@{}l@{}}R1. Recognize the daily activities of an older adult and detect abnormal patterns\\       - Waking up, Sleeping, Eating, SpareTime, Leaving, Toileting, Idle/Falling\end{tabular} \\ \midrule
F2. Analyze the status of an older adult through conversations &
  \begin{tabular}[c]{@{}l@{}}R2. Provide a communication method between a caregiver and an older adult \\        to assess the status of an older adult through a personalized question\end{tabular} \\ \midrule
F3. Preserve the independence and dignity of an older adult &
  \begin{tabular}[c]{@{}l@{}}R3-1. Reduce the usage of a camera and prioritize non-visual sensors to monitor activities\\R3-2. Allow older adults to control the system (e.g. what information to share)\end{tabular} \\ \bottomrule
\end{tabular}%
}
\end{table}

\subsection{Use Cases}\label{sect:system-usecases}
%After analyzing activities of daily livings from the dataset (Section \ref{sect:system-sensors}) and reviewing the literature, 
Based on the findings from our focus group session (Section \ref{sect:focusgroup-practices}), we specified five use cases of abnormal events. These use cases include 1) too frequent toilet usage, 2) abnormally leaving, 3) abnormal sleeping, 4) being idle for a long time, and 5) abnormal eating. One serious and prevailing disease for older women is urinary tract infections, in which an older adult might have urgency and frequent urination \cite{foxman2002epidemiology}. To detect these symptoms of urinary tract infections earlier, we included the first use case of frequent toilet usage. Also, we included abnormal leaving and sleeping, which are symptoms of dementia \cite{kales2015assessment}. Another widespread concern of older adults is falls, but there are no effective preventive solutions for falling \cite{wu2018understanding}. We included an abnormal event of being idle for a long time to detect a potential situation of falling. Lastly, as aging influences people's eating habits and affects their quality of life \cite{mond2005assessing}, we included the usage of detecting an abnormal eating event. 

Our five use cases based on the focus group are not an exhaustive list of use cases for an intelligent system \red{that monitors activities of daily living} for older adult care. Instead, the goal of these use cases is to demonstrate the functionalities of our system and receive the opinions of caregivers and older adults about the system. During the evaluation study, we also asked caregivers and older adults about additional use cases \red{that should be considered} (Appendix \ref{app_usecases})

\section{Prototype Implementation}\label{sect:system}
\begin{figure*}[t]
\centering
  \includegraphics[width=0.6\columnwidth]{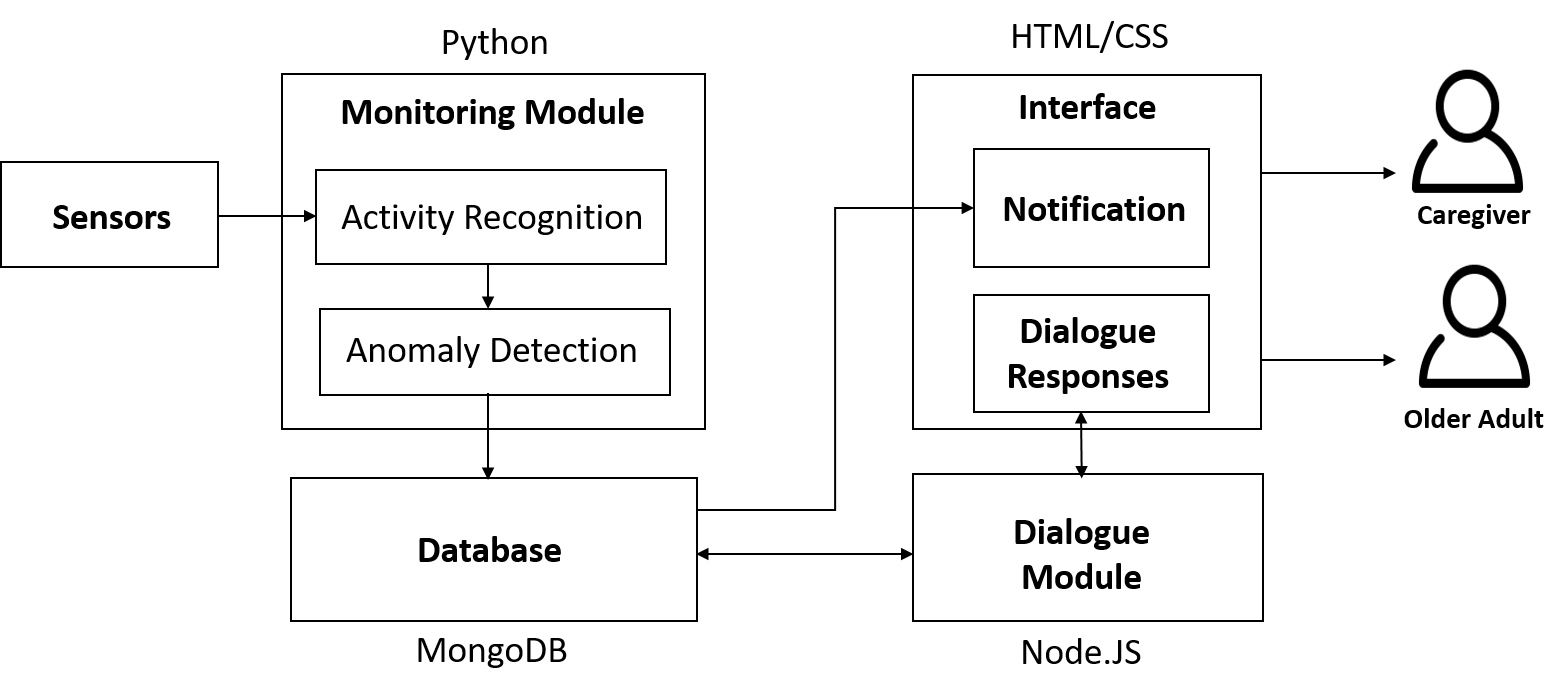}
  %\caption{High-Level Flow Diagram}
\caption{High-level flow diagram of an intelligent system for older adult care}\label{fig:flow-diagram}
\end{figure*}

Our main goal is to seek opinions and design ideas of caregivers and older adults on an intelligent system for older adult care. To \red{facilitate the elicitation of their opinions and ideas}, we developed an intelligent system to demonstrate its functionalities to caregivers and older adults. This system monitors the activities of an older adult and detects abnormal events using wireless motion sensors and machine learning. Also, this system utilizes interactive dialogue responses to explain abnormal patterns and allow caregivers to seek additional information about an older adult while an older adult can proactively determine what information to be shared. The  system consists of four major components: (a) monitoring module, (b) dialogue module, (c) interface of notifications and dialogue responses, and (d) database (Figure \ref{fig:flow-diagram}).

\begin{figure*}[t]
\centering 
\begin{subfigure}[t]{0.2\textwidth}
\centering
  \includegraphics[width=1.0\columnwidth]{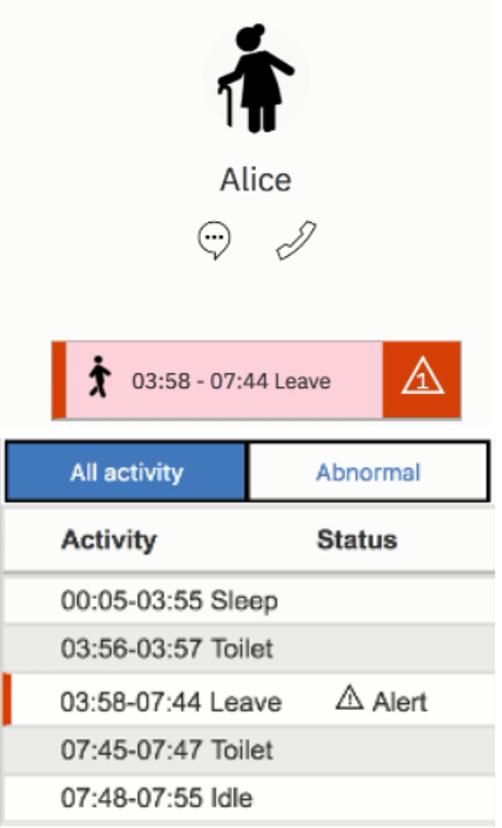}
  \caption{}
  \label{fig:system-int-noti}
\end{subfigure}\hspace{5mm}
\begin{subfigure}[t]{0.4\textwidth}
\centering
  \includegraphics[width=1.0\columnwidth]{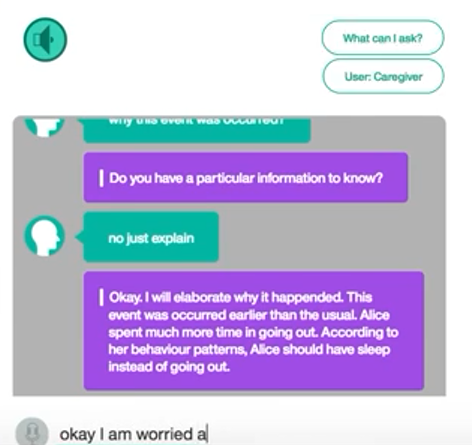}
  \caption{}
  \label{fig:system-int-diag-explain}
\end{subfigure}%\vspace{5mm}
\begin{subfigure}[t]{0.35\textwidth}
  \centering
  \includegraphics[width=1.0\columnwidth]{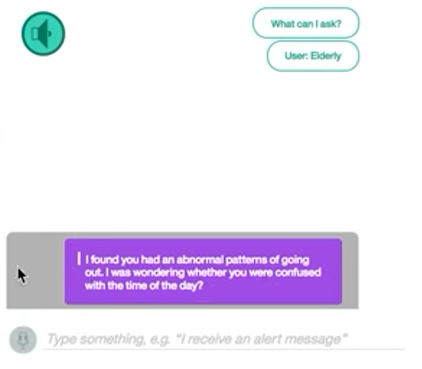}
  \caption{}
  \label{fig:system-int-diag-follow}
\end{subfigure}
\caption{Interface of the System: (a) Notification about an abnormal event, (b) Dialogue responses to explain an abnormal event for caregivers, and (c) Dialogue responses for older adults, in which they will receive a follow-up question by the caregiver and determine what information to be shared.}\label{fig:system-int}
\end{figure*}

For monitoring the daily activities of an older adult, this work focused on exploring wireless motion sensors instead of a camera to preserve the privacy of an older adult (Table \ref{tab:findings-requirements}) \cite{czaja2013older}. This work assumes that wireless motion sensors have been deployed in the house of an older adult or a care center. Given sensor data based on the activities of an older adult, the monitoring module will first recognize the activity of daily living and then detect whether a recognized activity pattern is abnormal or not. If the monitoring module detects any abnormal event, the system will send a notification to a caregiver on the web-based interface \red{(Figure \ref{fig:system-int-noti})}. As the abnormal notification does not provide sufficient information on the status of an older adult (Table \ref{tab:findings-requirements}), our system can explain the contextual information of an abnormal event to a caregiver \red{(Figure \ref{fig:system-int-diag-explain})}. If this explanation is still not sufficient, a caregiver can request our system to elicit any additional information to confirm the status of an older adult. Then, the system recognizes the activities of an older adult to determine an adequate moment for the interaction (e.g. taking a rest or being idle) and prompt dialogue responses (Figure \ref{fig:system-int-diag-follow}) to seek out the requested information from a caregiver (Table \ref{tab:findings-requirements}). Older adults can proactively control what information will be further shared through dialogue responses (Figure \ref{fig:system-int-diag-follow}). 
%All sensor readings indicate the presence in a specific area (e.g. sitting on a couch or lying in a bed) or occurrence in a particular event (e.g. cupboard open or toilet being flushed) by a single inhabitant. 

\subsection{\red{Sensor Data and Monitoring Module}}
For the development of a monitoring module, we utilized the dataset \cite{ordonez2013activity} that includes the recordings for 21 days of activities using 12 wireless \red{motion} sensors with the labels of 11 activities of daily living (ADL). We selected this dataset as it includes major daily activities that most caregivers leverage to understand the status of an older adult (Table \ref{tab:findings-requirements}). The dataset includes records of mounted wireless, infrared sensors in the inhabitant's house and track the daily activities of an inhabitant (a 57 year old male) in a specific area. Specifically, the sensor readings indicate the presence of a single inhabitant in a specific area (e.g. sitting on a couch) or the occurrence of a particular event (e.g. toilet being flushed). These measurements of the sensor values \red{were} discretized into time granularity $\Delta t$ = 60 seconds, which \red{was} experimented with as the best time slice from \cite{van2011human}. \red{Overall, the dataset included 30,240 time slices of monitoring ADLs.}

We referred \cite{ordonez2013activity} and \cite{shin2011detection} to implement our activity recognition and anomaly detection respectively.
For activity recognition, we \red{utilized} the Hidden Markov Model (HMM) \cite{rabiner1989tutorial,ordonez2014home} by learning the parameters using the maximum likelihood \cite{rabiner1989tutorial}. The observation, \red{feature} variable of the HMM is the vector of sensor readings ($x = {\vec{x_{1}}, \vec{x_{2}}, ..., \vec{x_{T}}}$) and the hidden variable is the activity label ($y = {\vec{y_{1}}, \vec{y_{2}}, ..., \vec{y_{T}}}$) for a total $T$ time slices. For the evaluation of activity recognition, we applied the leave-one-day-out cross-validation, in which we trained a model with data except data from one day for testing and achieved 0.85 accuracy and 0.62 F1-score. 

For anomaly detection, we \red{utilized} the following four contextual features of an activity: (1) the transition, (2) the duration, (3) frequency, and (4) starting hour of activity, which can be clues for caregivers to understand abnormal events (Section \ref{sect:focusgroup-practices}). For the transition of activity, we utilized the learned transition probability of the HMM as the prior of the transition of an activity. We then computed the posterior to estimate the likelihood of transition of an activity label. For other contextual features of an activity (i.e. duration, frequency, starting hour), we assumed that they follow Gaussian distribution and utilized the 90\% confidence intervals of corresponding Gaussian prior distribution to label an abnormal event \cite{shin2011detection}.
For the contextual feature of the transition, we indicated the transition is abnormal if the transition probability of activity is below 0.05. For the evaluation of anomaly detection, we applied leave-one-day-out cross-validation and explored traditional supervised learning algorithms (e.g. decision tree, naive bayes, and support vector machine). We selected a decision tree as our model, which has not only high performance (i.e. an average of 0.89 accuracy and 0.88 F1-score) but also model interpretability \cite{rudin2019stop}.

\red{As the focus of this work is to develop a prototype and receive early feedback on the functionalities of the system from older adults and caregivers}, this work does not explore other complex algorithms \cite{ordonez2016deep,ordonez2016deep} and sensor data to improve the performance of the monitoring module.

\subsection{Dialogue Module}
The dialogue module is designed and implemented to support the interactions with both caregivers and older adults (Table \ref{tab:findings-requirements}). Specifically, our dialogue model focuses on (1) generating the explanation of a detected abnormal event and (2) eliciting additional information or confirming the status of an older adult while older adults can proactively determine what information to be shared.

Our interface supports both chat-based interactions and voice-based dialogue responses using the Watson Cloud Speech-to-Text (STT) and Text-to-Speech (TTS) services. Given a user response, we utilized IBM Watson conversation service \cite{ibmconv} to recognize the intents or entities of the user response using at least 5 similar representative words. \red{Given the recognized intents or entities, we designed and implemented the state machine to generate audio and text outputs (Appendix - Figure \ref{fig:system-dialog-state-machine}). Specifically, if the recognized intent belongs to a specific state, our dialogue module will make a transition to a corresponding node of the state machine and generate audio and text outputs based on its specified template (Appendix \ref{app-dialogue}).}

\subsection{User Interface}\label{sect:system-int}
\red{We implemented the user interface using HTML and CSS for (1) providing the notification of an abnormal event and (2) supporting dialogue responses with a caregiver or an older adult.} 

The notification interface includes only brief information about an abnormal event (e.g. which abnormal activity is detected and when it has occurred) (Figure \ref{fig:system-int-noti}). To highlight the detected abnormal event, we included red background color. The dialogue response interface aims to explain an abnormal event to a caregiver (Figure \ref{fig:system-int-diag-explain}) and elicit any additional information or status of an older adult (Figure \ref{fig:system-int-diag-follow}). In addition, older adults can proactively control what information to be shared through the dialogue response interface (Figure \ref{fig:system-int-diag-follow}). A user can type a text response or use the voice recognition function by clicking a grey button at the bottom left of Figure \ref{fig:system-int-diag-explain} and \ref{fig:system-int-diag-follow}.

\subsection{Database}
We implemented a Publish-Subscribe database with MongoDB to store \red{processed} data \red{of our system} and built a communication pipeline among the modules of our system. For example, after recognizing activities of daily living and detecting their abnormalities, the monitoring module will store its output and relevant data in the database. If an abnormal event is detected, the database will send the corresponding outputs to the notification interface or the dialogue module to support user interactions.

\section{Qualitative Study}
We conducted semi-structured interviews to understand the opinions of professional caregivers and older adults about the designs and functionalities of our intelligent system (Section \ref{sect:system}) for older adult care.

\subsection{Participants}
For the qualitative study, we recruited five professional caregivers (C5 - C9 in Table \ref{tab:participants-caregivers-eval}) and five older adults (O1 - O5 in Table \ref{tab:participants-olderadults}) through word-of-mouth and sharing the flyer to local senior centers. 
Professional caregivers (C5 - C9 in Table \ref{tab:participants-caregivers-eval}) have 1 - 32 years of experience (M = 17.8, SD = 13.4) in assisting older adults at a daycare center. Their care activities include, but are not limited to, taking older adults (with/without dementia and mobility limitations) to the toilet, feeding, hygiene assistance (e.g. bathing, showering), and assisting with anything that is part of older adults' daily lives (e.g. dressing clothes). 
%%%%
%Their ages ranged from 20 - 59 (M = 47.6, SD = 14.2). Their scores on experience with technology ranged from 2.3 - 6.8 out of 7.0 (M = 3.6, SD = 1.8). Except for C4 and C5, other caregivers (C1 - C3) have limited experience with technology (below a score of 3.5). 

For the recruitment of older adults, we excluded participants with severe cognitive or communication impairments to participate in the interview. Unfortunately, we were unable to recruit an older adult living alone or living at a center. 
Older adults shared their age, gender, and experience of caregiving if exists. In addition, they filled out questionnaires \cite{czaja2006factors,beer2012domesticated} to assess their experience with recent technologies (e.g. smartphones, computers, AI services). In addition, older adults provided additional information on their living conditions and health issues.  
Recruited older adults' ages ranged from 60 - 79 (M = 66.6, SD = 7.4). Among five older adults, O4 and O5 have experience in caregiving to family members. Older adults' scores on experience with recent technologies ranged from 2.3 - 4.0 out of 7.0 (M = 3.1, SD = 0.7), where a score of 7 indicates high proficiency in technologies. Overall, most older adults have limited proficiency in technologies (below a score of 3.5) except for O1, who has decent experience with technologies with a score of 4.0.

% \usepackage{graphicx}

\begin{table}[htp]
\centering
\caption{Demographics information of professional caregivers for the qualitative study}
\label{tab:participants-caregivers-eval}
\resizebox{0.8\textwidth}{!}{%
\begin{tabular}{cccl} \toprule
\textbf{PID} & \textbf{Studies} &
  \textbf{Gender} &
  \textbf{Caregiving Experience} \\ \midrule
C5 & Evaluation \& Design &
  Female &
  \begin{tabular}[c]{@{}c@{}}30 years of professional caregiving at a day care center\end{tabular} \\
C6 & Evaluation \& Design &
  Female &
  \begin{tabular}[c]{@{}c@{}}32 years of professional caregiving at a day care center\end{tabular}  \\
C7 & Evaluation \& Design &
  Female &
  \begin{tabular}[c]{@{}c@{}}8 years of professional caregiving at a day care center\end{tabular}  \\
C8 & Evaluation \& Design &
  Female &
  \begin{tabular}[c]{@{}c@{}}18 years of professional caregiving at a day care center\end{tabular} \\
C9 & Evaluation \& Design &
  Female &
  \begin{tabular}[c]{@{}c@{}}1 year of professional caregiving at a day care center\end{tabular} \\ \bottomrule
\end{tabular}%
}
\end{table}

\begin{table}[htp]
\centering
\caption{Demographics information of older adults for the qualitative study}
\label{tab:participants-olderadults}
\resizebox{\textwidth}{!}{%
\begin{tabular}{ccccccc} \toprule
\textbf{PID} &
  \textbf{\begin{tabular}[c]{@{}c@{}}Age\\ Range\end{tabular}} &
  \textbf{Gender} &
  \textbf{Caregiving Experience} &
  \textbf{\begin{tabular}[c]{@{}c@{}}Technology\\ Experience\end{tabular}} &
  \textbf{Living Conditions} &
  \textbf{Health Issues} \\ \midrule
O1 &
  60 - 69 &
  Female &
  N/A &
  4.0 out of 7.0 &
  With families &
  \begin{tabular}[c]{@{}c@{}}Diabetes, \\ High blood pressure, \\ and Gallstones\end{tabular} \\
O2 &
  70 - 79 &
  Female &
  N/A &
  2.3 out of 7.0 &
  With the married partner &
  \begin{tabular}[c]{@{}c@{}}Diabetes, \\ High blood pressure, \\ and Cardiac Insufficiency\end{tabular} \\
O3 &
  70 - 79 &
  Male &
  N/A &
  2.8 out of 7.0 &
  With the married partner &
  \begin{tabular}[c]{@{}c@{}}Diabetes, \\ High blood pressure, \\ and Rheumatoid arthritis\end{tabular} \\
O4 &
  60 - 69 &
  Female &
  \begin{tabular}[c]{@{}c@{}}5 years of caregiving to families \\ (e.g. mother-in-law and aunt)\end{tabular} &
  3.8 out of 7.0 &
  With the married partner &
  \begin{tabular}[c]{@{}c@{}}Dental Issues\\ and Varicose veins\end{tabular} \\
O5 &
  60 - 69 &
  Male &
  \begin{tabular}[c]{@{}c@{}}1 years of caregiving to families \\ (e.g. father and mother-in-law)\end{tabular} &
  2.5 out of 7.0 &
  With the married partner &
  \begin{tabular}[c]{@{}c@{}}Thyroid cancer\\ and Diabetes\end{tabular} \\ \bottomrule
\end{tabular}%
}
\end{table}

\subsection{Procedure}
We conducted the remote interviews over a video conference platform (i.e. Zoom). In each interview, the researcher first introduced the project and the functionalities and components of the system (e.g. wireless motion sensors, data analysis, and interfaces to send an alert notification and support interactive dialogue responses). In addition, inspired by the previous design works on older care technologies \cite{beer2012domesticated,lee2022enabling}, we showed the video demonstration of our system with the use case of an abnormal toilet event (Section \ref{sect:system-usecases}) to illustrate the system functionalities and possible interactions with it. Note that the system was not deployed. For the sake of time, we only showed only one use case on an abnormal toilet event and just enumerated other use cases by words. The utilized video demonstration of our system can be found in the following link: [anonymized].
%the supplementary material or the following anonymized link. 
%\textbf{\url{shorturl.at/asAG0}}

After introducing our project and system functionalities with the video demonstration, we inquired about a series of questions relating to using an intelligent system that monitors and detects abnormal events of an older adult and provides interactive dialogue responses to control what information to be elicited or shared. Specifically, we asked participants to describe 1) new use cases of the system, 2) the potential benefits and limitations of the system, 3) whether they can have a \red{trustworthy} usage of the system, 4) their willingness to interact with the system, and 5) the privacy issue of the system. Finally, participants were asked to envision new designs and ideas that could improve the system and the care practice of an older adult. 
%Finally, we conducted a design activity, in which participants were asked to envision new designs and ideas that could improve the system and the care practice of an older adult. 
Although the video served as a foundation to discuss the usability aspects of the system, we clarified to the participants that new designs and ideas were not limited to what was shown in the video. The caregivers and older adults provided their comments from the perspective of a caregiver and an older adult being monitored respectively. Each interview took around an hour and the procedure was approved by the Ethical Committee.

\subsection{Analysis}
All interviews were audio-recorded and transcribed for further analysis. For the data analysis, we followed an interactive, inductive thematic analysis \cite{braun2006using}. First, we generated initial codes from the structured topics of the interviews. After reviewing transcripts, the researchers generated codes and findings and iteratively discussed the codes of the transcripts with the research team. 
%We describe the details in the next section. 

%We describe the overall thematic analysis in () and describe the details in the next section. 

\section{Findings of Qualitative Study}
Both professional caregivers and older adults appreciated our intelligent system that detects and explains an abnormal event of an older adult and further empowers older adults and caregivers to communicate with each other and control the system for personalized care services. Especially, they showed positive opinions on our system which has the potential to provide faster attentive and personalized care services, reduce cognitive burdens and workloads of caregivers, and improve social interactions of older adults. In addition, they valued the functionalities of our system to explain the contexts of an abnormal event and elicit additional information while older adults can proactively decide what to be shared for more empowering and trustworthy usage of the system. They were willing to interact with our system which utilizes wireless motion sensors to preserve the privacy of an older adult. However, they also expressed practical concerns and challenges, such as questionable system performance and efficiency, dependency on the complex system, limited ways to interact, and financial and policy aspects. Furthermore, they provided new design ideas to enhance use cases and functionalities of the system adult care for more empowering and \red{trustworthy} interactions with the system. 
In the next section, we describe the detailed findings of our qualitative evaluation of our system: opportunities \& challenges of the system (Table \ref{tab:finding-procons-trustprivacy}), trust \& privacy aspects of the system (Table \ref{tab:finding-procons-trustprivacy}), expected intention to interact with the system, and additional use cases and functionalities (Section \ref{sect:finding-usecases}).
%details in Appendix \ref{app_usecases}). 

\begin{table}[h]
\centering
\caption{Summarized findings on (1) the potential opportunities \& challenges and (2) trust \& privacy aspects of the intelligent system that monitors and detects an abnormal event of an older adult}
\label{tab:finding-procons-trustprivacy}
\resizebox{0.85\textwidth}{!}{%
\begin{tabular}{ll}\toprule
\multicolumn{1}{c}{\textbf{Perceived Opportunities}} &
  \multicolumn{1}{c}{\textbf{Perceived Challenges}} \\ \midrule
\begin{tabular}[c]{@{}l@{}}1. Faster Attentive and Personalized Care Services\\ 2. Reduce the Cognitive Burdens and Workloads of Caregivers\\ 3. Improve the Control \& Social Interactions of Older Adults\end{tabular} &
  \begin{tabular}[c]{@{}l@{}}1. Questionable System Performance and Efficiency\\ 2. Dependency on a Complex System\\ 3. Limited Ways to Interact\\ 4. Financial and Policy Aspects \& Disparities in Services\end{tabular}\\ \toprule
\multicolumn{2}{c}{\textbf{Trust \& Privacy Aspects of Using the System}}\\ \midrule 
\multicolumn{2}{c}{\begin{tabular}[c]{@{}l@{}}1. Convincing with the System Performance \& Functionalities\\ 2. Explain and Control What to Share\end{tabular}}\\ \bottomrule
\end{tabular}
}
\end{table}

\subsection{Perceived Opportunities and Usefulness of the System}\label{sect:finding-opportunities}

\subsubsection{Faster Attentive and Personalized Care Services}\label{sect:finding-opp-fast}
%\hfill
%
%\noindent
One potential benefit that caregivers and older adults appreciated the most is the possibility of providing faster attentive and personalized care services for an older adult. Caregivers mentioned that it is
%\begin{quote}
\textit{``not possible to have an employee in each room''} (C9). \textit{``This is a new reality (…) it alerts us to certain situations that are out of the ordinary so that there is a faster service in responding to that problem (...) This will allow personaliz\red{[ing]} your attention''} (C5). %\textit{``The fact that this system allows you creating an appropriate record for each person according to their needs/status from interactions, that really is a plus''} (C5)
%\end{quote}

In addition to caregivers' comments, older adults shared their anecdotes that describe the potential value of our system to provide prompt care services. 
%
%\begin{quote}
%\textit{``When people are alone, sometimes things happen that we don't even dream of''} (O3).
\textit{``a lot of older adults who live alone and get unnoticed about their abnormality''} (O1). \textit{``if we fall or suddenly lose sight or lose the sense of what we are doing, this system can be good to receive clinical assistance immediately''} (O2). \textit{``If I have such a system earlier, (...) I could have detected the falling accident of my mother-in-law much earlier and provided timely care.''} (O4).
%\end{quote}

\subsubsection{Reduce the Cognitive Burdens and Workloads of Caregivers}\label{sect:finding-opp-burdens}
%\hfill
%
%\noindent
Another perceived usefulness of our system is to reduce the workloads of caregivers. As caregivers typically have a number of older adults to provide care services, they often have cognitive burdens to keep track of what each older adult needs. They \textit{``sometimes forget things for an older adult so that it would be helpful in that respect to receive an alert notification about the  situation''} (C9).

In addition, most care services require visiting each older adult, manually tracking his/her status, and providing assistance if necessary. Both caregivers and older adults considered that the abnormal detection of our system has the potential value to \textit{``reduce the burden of caregivers''} (O5) and improve care services by having \textit{``more time to pay attention and spend with older adults''} (C7)

\subsubsection{Improve the control \& social interactions of older adults}\label{sect:finding-opp-social}
%\hfill
%
%\noindent
%Older adults described that our system has the potential value of improving the social interactions of an older adult. 
Older adults appreciated the value of interactive dialogue responses to improve the control of the system and counteract the isolation of an older adult. 
%Similar to the previous studies that discussed the pressing issues of privacy invasion and the lack of controls of older adult care technologies \cite{berridge2022control,nurgalieva2019information,vines2013making}, 
Even if older adults appreciated the benefits of older adult care technologies, they were \textit{``reluctant to being always tracked and asked what they do''} (O5). However, older adults found that our system concepts enable them to have more autonomy and control of the system: \textit{``This system is nice because (...) this won’t control our life and activities directly. Instead, it will preserve our independence''} (O2). This system is considered \textit{``better as it [the system] shares and explains the minimal information of abnormal situations, I can determine what further information to be shared''} (O4). %\textit{``I 

As individuals age, their social relationships may change for various reasons 
%(e.g. migration of family members, a decline in physical or cognitive abilities)
\cite{cudjoe2020epidemiology}. Social isolation is a prevalent problem of older adults, which might lead to numerous detrimental health conditions \cite{nicholson2012review, cudjoe2020epidemiology}. Similar to the prior research work on a companion agent \cite{sidner2018creating}, older adults considered that the value of interactive dialogue responses to reduce the feeling of social isolation: \textit{``even having a conversation with a system to check the status would help reduce the feeling of isolation from an older adult''} (O1).

\subsection{Perceived Limitations \& Challenges of Using the System}\label{sect:finding-limitations}

\subsubsection{Questionable System Performance and Efficiency}\label{sect:finding-limit-systemperf}
%\hfill
%
%\noindent
Both caregivers and older adults expressed concerns about the system performance. For instance, caregivers questioned whether a system can perform well with the presence of multiple older adults: \textit{``many older adults in the care center (...) it would be difficult to have a system that would monitor all the visits to the bathroom of each one''} (C9). In addition, caregivers and older adults are concerned whether a system can monitor detailed activities (e.g. medication taking): \textit{``more difficult to understand if an older adult really takes them [medication] or not''} (C6).

\subsubsection{Dependency on a Complex System}\label{sect:finding-limit-complex}
%\textbf{Technical Dependency}\\
%\textbf{Not accustomed to recent technologies}\\
%\hfill
%
%\noindent
Caregivers had concerns about having constraints on their workflows \textit{``if we (they) have to be dependent on being connected to a system''} (C5). In addition, as \textit{``older adults are not used to technologies and they do not know what it consists of, how it going to work well or not''} (C9), caregivers are not sure how well an older adult can successfully use a system.

\subsubsection{Limited ways to interact}\label{sect:finding-limit-interact}
%\hfill
%
%\noindent
Older adults described the limitation of interactive dialogue responses to interact and notify an urgent situation. Specifically, older adults questioned whether they have to be \textit{``at a certain location to engage interactive dialogue responses with the system?''} (O5). In addition, as interactive dialogue responses require multiple steps to confirm the status of an older adult and seek assistance, they are \textit{``worried about the situation when an older adult cannot respond to the system through dialogue responses (...) Instead of waiting a few minutes or hours (...) it is critical to provide adequate care immediately when a person is under emergency''} (O4).

\subsubsection{Financial and Policy Aspects \& Disparities of Services}\label{sect:finding-limit-financial}
%\hfill
%
%\noindent
Caregivers and older adults described the practical concerns on financial and policy aspects of using a system. Caregivers mentioned that \textit{``We are positive to explore such a technology''} (C5 and C6). \textit{``However, the main problem is the lack of financial investment (…) there would have to be some partnership with the state, or whatever''} (C9).

In addition, older adults elaborated on the possible disparity in care services due to an intelligent system: \textit{``I think such technology-based care assistance might lead to unequal distributions of care support. There could be an older adult, who might not be able to afford these systems and (...) excluded by receiving such remote cares''} (O5).

\subsection{Trust \& Privacy Aspects of Using the System}\label{sect:finding-trust-privacy}
\subsubsection{Convincing with the System Performance and Functionalities}\label{sect:finding-trust-describe}
%\hfill
%
%\noindent
Both caregivers and older adults desire to know about the system performance and functionalities to have empowering and \red{trustworthy} usage of a system. For instance, caregivers described that conducting in-field validation and describing how well a system can perform its functionalities would help support more \red{trustworthy} usage of a system. Caregivers are uncertain whether an older adult can have the \red{trustworthy} usage of a system, because \textit{``anything new to older adults is strange and they are suspicious of everything''} (C8).
According to the interviews with older adults, we found that they do not particularly seek system performance. Instead, they described that it is important to demonstrate and convince with \textit{``the functionalities of the system that provides a new value for us \red{[older adults]}.''} (O2). They are
\textit{``willing to trust and use this system because at least the purpose of this system is to protect any abnormal events that happen to us''} (O3). \textit{``If I had my parents, I would like to trust and use it [the system], which can help me find out how older adults are doing and have them some companionship''} (O1).

\subsubsection{Explain and Control What to Share}\label{sect:finding-explain-control}
Both caregivers and older adults appreciated our system that leverages wireless sensors to preserve the privacy of an older adult: \textit{``Since the system operates by sensors and not a camera and video recordings, I don't think there would be much of a problem. However, (…) I don't know to what extent they [older adults] would accept this happening''} (C9).

Our interviews showed that older adults were mostly \textit{``okay with tracking and sharing my presence and activities at specific time and location in the house''} (O3). Older adults found that utilizing wireless motion sensor data instead of videos from a camera will be \textit{``more likely to preserve the privacy of an older adult''} (O4). \textit{``As dangerous situations can happen anywhere in a house, it is okay to track the presence at the home. (...) I do not find a particular privacy concern as this system is made of wireless sensors''} (O1). However, they described that it is also \textit{``important to explain what information will be shared and let an individual determine how much data an older adult is willing to share''} (O5).

Older adults appreciated our system that detects and explains an abnormal event and provides means of interacting and controlling a system through interactive dialogue responses instead of having a fully automated approach: 
%\begin{quote}
\textit{``If the system just notifies the occurrence of an abnormal event, we might be confused in what contexts abnormality occurs (...) I like the concept of a system that provides additional contexts [(e.g. factors of abnormal detection – duration, frequency)]. (...) good to have interactive dialogues responses from the system''} (O4), which \textit{``empowering me[an older adult] to share what information to be shared''} (O5).

\subsection{Expected Intention to Interact with the System}\label{sect:finding-interaction}
Caregivers expressed concern on whether older adults can interact with the system, because \textit{``Older adults have not grown up in this age of computers, cell phones, etc''} (C9). In contrast to caregivers' concerns, we found that all older adults were positive to interact with our system that provides several opportunities to improve care services (Section \ref{sect:finding-opportunities}): older adults were \textit{``willing to engage in interactions with the system to clarify my situation even if I do not have the necessary smartphone to interact with the system''} (O2).

Overall, we found that older adults were \textit{``okay with interacting with the system through voice or typing as a message to the system''} (O3). However, as \textit{``an older adult might not very familiar with having chats or conversations with an intelligent system''} (O5), older adults desire some \textit{``training''} (O2) or \textit{``tutorials''} (O5) about the system to interact with the system. In addition, older adults desired more diverse ways to interact with the system based on the context of an older adult (Section \ref{sect:finding-limit-interact}) and the physical status: \textit{``As each older adult might have different visual and auditory status, the system interface should be provided various interaction methods accordingly''} (O4).

\subsection{Additional Use Cases \& Functionalities}\label{sect:finding-usecases}
We summarize additional use cases and functionalities that professional caregivers and older adults suggested as follows:
(1) monitoring other fine-level daily activities, (2) environmental monitoring (e.g. the presence of others \& temperature), (3) physical \& cognitive assistance, (4) validation of medical symptoms, and (5) improving communications among stakeholders (e.g. professional caregivers, family members, doctors). The detailed comments on each use case \& functionality by caregivers and older adults can be found in Appendix \ref{app_usecases}.

\section{Towards More Empowering Older Adult Care Technologies}\label{sect:finding-designs}
In this section, we discuss design considerations for a more empowering and controllable intelligent system for older adult care. 
Throughout our qualitative study with an design activity, professional caregivers and older adults discussed their perceived opportunities (Section \ref{sect:finding-opportunities}) and limitations (Section \ref{sect:finding-limitations}) of an intelligent system for older adult care. 
Specifically, caregivers appreciated the potential of our system functionalities to promptly check the status of an older adult and elicit further missing information from an older adult through dialogue responses. Older adults mentioned the value of interactive dialogue responses to proactively determine what information to be shared. 
In addition, caregivers and older adults provided other perspectives on ways to support more empowering and \red{trustworthy} user interactions with the system (Section \ref{sect:finding-trust-privacy}). This work discusses engagement with users as one of the driving forces to make human-centered, older care technologies along with other considerations%recommendations
: (1) explainability \& interactivity, (2) new benchmark datasets \& human-centered evaluations, (3) multimodal sensing \& omnipresent interactions, and (4) intuitive designs \& tutorials.

%\subsection{Ways for Empowering \& {Trustworthy} Older Adult Care Technologies}\label{sect:finding-designs-trustfulengagement}

%\subsubsection{Explainability \& Interactivity}\label{sect:finding-designs-trustfulengagement-xai}
\subsection{Explainability \& Interactivity}\label{sect:finding-designs-trustfulengagement-xai}
Our work discusses the importance of making older adult care technology explainable and interactive \cite{holzinger2017we,gunning2019xai} for more controllable and trustworthy user interactions. Caregivers appreciated the potential of our system functionalities to promptly check the status of an older adult and determine if an intervention is necessary. As older adult care technologies with machine learning models cannot be perfect and do not provide the whole picture of the status of an older adult, our system provides explanations on the status of an older adult with contextual information (e.g. the transition, frequency, duration, and time of activities). When the information from the system is not sufficient to understand the status of an older adult, caregivers can utilize interactive dialogue responses to elicit further missing information from an older adult for personalized older adult care services. 

Older adults described that they \textit{``do not have any particular numerical values on how accurately a system makes a prediction. Instead, what is the most important for me to trust is how well a system can be adapted to my personal health situation and life patterns and provide personalized interactions''} (O4). When our system detects an abnormal event, our system explains the abnormal event with contextual information to a caregiver and allows older adults to proactively share their information through interactive dialogue responses. As our system provides \textit{``a means of communicating to confirm the situation and derive a personalized interaction, I [an older adult] would be able to trust this system more''} (O4). Our study showed that explanations and interactive dialogue responses have the potential for personalization and described other various aspects to realize personalization: 
\textit{``providing a way to inform and control what notification will be given and how it will be given (...) such as opt-out monitoring a certain situation''} (O4), and adaptive usage of multimodal sensor data to preserve the privacy of an older adult (Section \ref{sect:finding-designs-trustfulengagement-multi}).

In this work, we contributed to the explainable older adult care technologies \cite{holzinger2017we,gunning2019xai,wolf2020designing} for detecting abnormal daily activity patterns of an older adult. Specifically, this work utilized an interpretable model (i.e. a decision tree) and translated sensor data and contextual features into natural language descriptions that are meaningful for the end-user. However, translations to generate natural language explanations are limited to wireless sensor data and activity of daily living that this work focused on. It is important to further explore how to design explainable older adult care technologies for other applications (e.g. detecting cognitive declines \cite{zubatiy2021empowering,pollack2005intelligent}), data modalities, and stakeholders with different purposes of interacting with the systems \cite{suresh2021beyond}.

%\subsubsection{New Benchmark Datasets \& Human-Centered Evaluations}\label{sect:finding-designs-datasets}
\subsection{New Benchmark Datasets \& Human-Centered Evaluations}\label{sect:finding-designs-datasets}
Our study implies that there is a gap between the curated datasets \cite{ordonez2013activity,ordonez2015sensor,nweke2018deep} for the system implementation and evaluation and diverse realistic situations mentioned by older adults and caregivers. Our study identified the necessity of further investigation to create new benchmark datasets and design evaluation methods for older adult care technologies. As most datasets include the recording of sensor data from a single inhabitant \cite{ordonez2013activity,ordonez2015sensor,nweke2018deep} and a few sets of daily activities, it remains unclear how a machine learning model can perform other unseen situations, such as the presence of multiple inhabitants, disruptions of sensor data based on a request by an older adult, and expansion to a new activity. For deploying older adult care technologies in practice, it is critical to understand the shortcoming of the current datasets and evaluation and explore how we can create datasets that support more diverse or failure situations or evaluation metrics and methods.

In addition, our study shed light on the possible disparities in older adult care services for people who cannot afford older adult care technologies. Instead of focusing on optimizing and evaluating older adult care technologies with respect to their system performance, our work also suggests the importance of investigating the ethical implications of adopting these older care technologies \cite{ho2020we}, how we can derive more equitable care services through technologies \cite{schwalbe2020artificial}, and human-centered evaluations of these technologies.

\subsection{Multimodal Sensing \& Omnipresent Interactions}\label{sect:finding-designs-trustfulengagement-multi}
This work discusses the need of exploring multimodal sensing and machine learning models \cite{perry2004multimodal,kumari2017increasing,ngiam2011multimodal} and omnipresent interactions for improving the system performance and experiences of older adult care technologies. Both caregivers and older adults have concerns about the performance and efficiency of the system (e.g. detecting an abnormal event) (Section \ref{sect:finding-limit-systemperf}). O5 commented that \textit{``If a system relies only on wireless motion sensors, it might be difficult to monitor various situations''} and recommended to \textit{``complement wireless motion sensors with images from cameras''}. As the usage of cameras might invade the privacy of an older adult, O5 further described to \textit{``leverage only the silhouette of an older adult from a camera or sensor data from a wrist band to predict the status or show it to a caregiver''} (O5). Thus, we recommend considering the potential of exploring multimodal data to improve the system performance on understanding fine-level daily activities while providing a means of controlling multimodal sensing in different situations (e.g. enabling silhouette sensing on high risks of a falling event). 

In addition to the exploitation of multimodality \cite{turk2014multimodal}, the investigation of omnipresent interactions would be also beneficial to improve user experience with the system. Although older adults appreciated the value of interactive dialogue responses to improve the control and personalized interactions of the system, they considered that dialogue responses of our system provide limited interactions during an emergency (Section \ref{sect:finding-limit-interact}). During the design activities, older adults elaborated on the necessity of a new interaction: \textit{``There are going to be multiple emergency scenarios and we need a way to notify an emergency situation accordingly: one where an older adult is unconscious and requires relying on sensors and proactive intervention of a caregiver. The other, where an older adult is conscious and can make some sounds or movements to notify an emergent situation''} (O5). \textit{``If an older adult falls with some consciousness and bleeding, but cannot move. In that case, it would be helpful if an older adult can make a sound of `sos' and a system detects this to request assistance''} (O4). We suggest investigating various, omnipresent interaction methods using multimodality for effectively coping with various situations.

%\subsubsection{Intuitive designs \& tutorials}
\subsection{Intuitive designs \& tutorials}
Both caregivers and older adults described the necessity of having an intelligent system with intuitive designs. Specifically, C5 mentioned that \textit{``there is a fixed team but generally new people are always coming and going. So it would have to be something that didn't require a lot of learning and that was intuitive and simple''} (C5). As older adults might not be familiar with recent technologies, \textit{``the system has to be easy to use''} (O5). It is important to iteratively engage with users and identify what they consider as intuitive designs and interactions of the system. 

Even if a system has user-centered and intuitive designs, we discuss that tutorials on a system would be beneficial to positively influence the acceptance and usage of a system. Prior work explored older adults' acceptance of mobile technology and described that the tutorial training significantly affected older adults' acceptance of mobile technologies \cite{wilkowska2009factors}. As older adults might not \textit{``know little or nothing about technologies''} (O2), we draw parallels with the prior work on older adults' acceptance of mobile technology \cite{wilkowska2009factors} and consider that \textit{``training to use the system might be needed''} (O2).

\section{Limitations}
We worked with a group of family and professional caregivers and older adults, who have relatively diverse characteristics in age range, caregiving experience, and technology experience. Including family and professional caregivers and older adults with diverse characteristics was helpful to have a holistic perspective of overall caregiving practices for older adults. However, we only recruited 14 participants (nine caregivers and five older adults) and did not get to recruit older adults, who have cognitive impairments due to their difficulty with participating in interviews, and older adults, who live in a center or alone. While our findings and design considerations are applicable to intelligent, older adult care technologies in general, additional studies with larger groups of caregivers and older adults with diverse living conditions, and physical and cognitive conditions would provide additional perspectives on these technologies. 

In addition, we acknowledge the limitation of our qualitative study that relied on the video demonstration of our system functionalities.
Although prior design studies on care technologies showed the value of showing video demonstrations \cite{beer2012domesticated,lee2022enabling}, caregivers and older adults might provide new perspectives when they directly interact with the system. Therefore, future work is required to conduct an evaluation study with caregivers and older adults through direct interactions and how to improve multimodal sensing and interactions for more empowering and trustworthy older adult care technologies.

\section{Conclusion}
This work contributed to designs of improving the control of older adult care technologies through engagement with family and professional caregivers and older adults. Specifically, we conducted the focus group session with family caregivers to specify the design scope of the system and then developed the high-fidelity system prototype. This system can detect and explain an abnormal event of an older adult using wireless sensor data and machine learning models and utilizes interactive dialogue responses that enable caregivers to elicit missing information from an older adult and older adults to proactively control what information to be shared.
Our qualitative study with older adults and professional caregivers showed that they appreciated the perceived opportunities of our system, which can provide faster attentive and personalized care.
They also described the value of our system with interactive dialogue responses that not only explains an abnormal event, but also elicits additional information about an older adult while allowing older adults to proactively determine what information to be shared for more trustworthy and personalized care services. In addition, we discuss design considerations, such as explainability \& interactivity, new benchmark dataset \& human-centered evaluations, multimodal sensing \& omnipresent interactions, and intuitive designs \& tutorials) to improve the control and adoption of older adult care technologies.


\bibliographystyle{ACM-Reference-Format}
\bibliography{main}

%\newpage
\appendix

\section{Prototype Implementation Details}

\subsection{Dialogue Interface}\label{app-dialogue}
\begin{figure*}[htp]
\centering 
  \includegraphics[width=0.4\columnwidth]{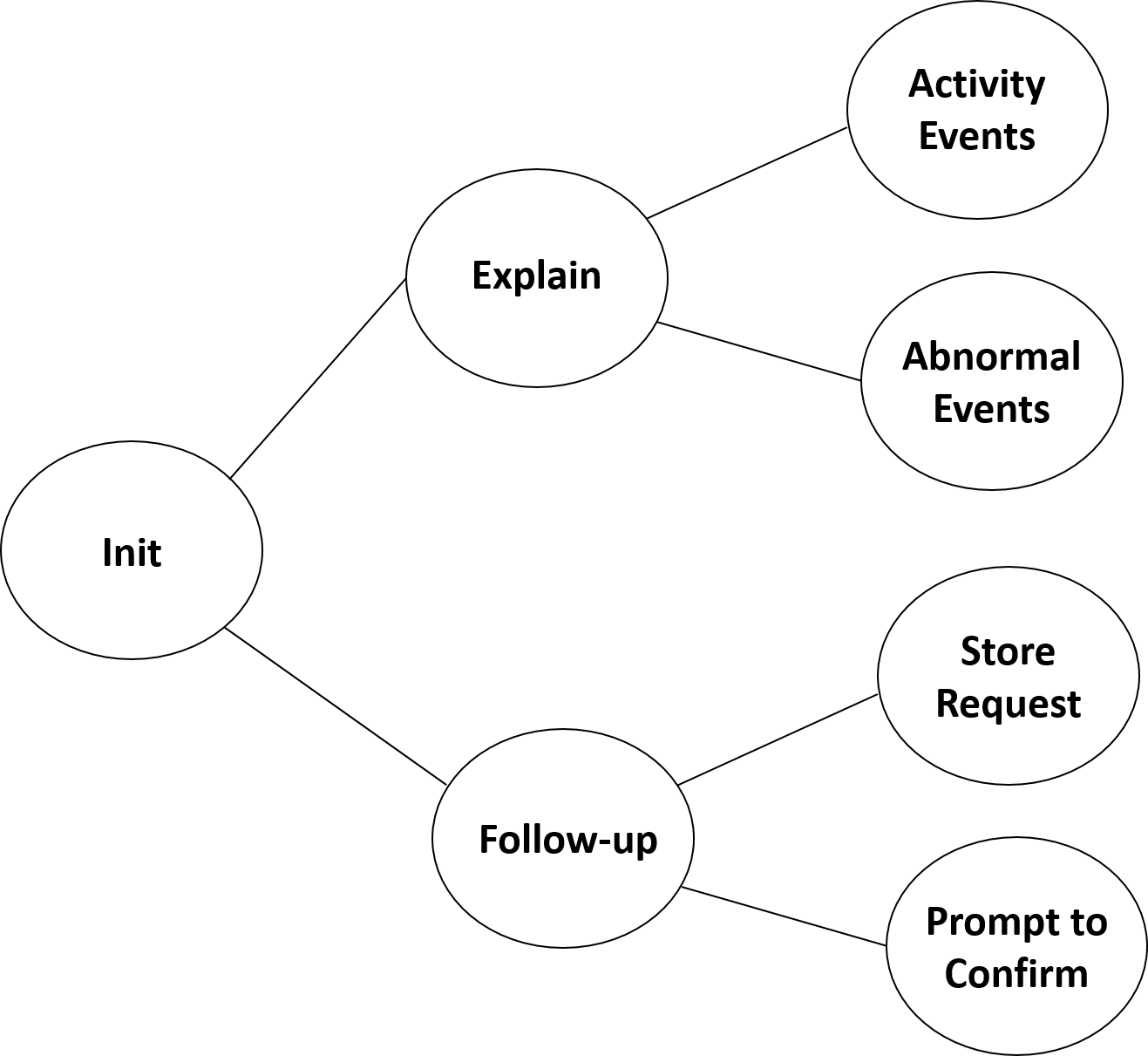}
\caption{State machine to generate dialogue responses: When a caregiver or an older adult initiates the interface, the system will generate the greeting message at the \textit{``Init''} state. If a user requests to explain a detected activity or abnormal event, contextual information (e.g. time of a recognized activity, contextual features that contribute to an abnormal event) at the \textit{``Explain''} state. In addition, when a caregiver requests to elicit missing information, the system will be at the \textit{``Store Request''} state and store a corresponding request. When an older adult either takes a rest or is being idle, the system will prompt a question/request to elicit additional information at the \textit{``Prompt to Confirm''} state.}\label{fig:system-dialog-state-machine}
\end{figure*}

For generating dialogue responses, we designed and implemented the state machine (Figure \ref{fig:system-dialog-state-machine}), which hierarchically represents major interaction functions into multiple child states. 

The \textit{``Init''} state generates the greeting dialogue to the user after retrieving the name of the user from the database. 

The\textit{``Explain''} state and its child states provide an explanation of the detected activity event. They can either describe the general information of an event or explain contextual features of abnormal events.  The \textit{``Activity Events''} state will utilize the extracted information from activity recognition. The template of dialogue generation is specified as follows: \textit{``@user @what @where @when''}, where \textit{``@user''} is the name of a recognized user, \textit{``@what''} represents a verb representation of recognized activity, \textit{``@where''} describes a location of recognized activity, \textit{``@when''} is a time of recognized activity. A sample output response is described as follows: \textit{``Mike took a rest in the living room at 8:30''}.

The \textit{``Abnormal Events''} state will check which contextual feature is detected as abnormal and describe why that factor is detected as abnormal. The template of dialogue generation is specified as follows: \textit{``@user @what @abnormal\_feature. + @user @prediction''}. \textit{``@user''} and \textit{``@what''} are defined the same as described earlier: the name of a recognized user and a verb representation of a recognized activity. \textit{``@abnormal\_feature''} represents an adjective form of a feature of an abnormal detector: the adjective form of a duration feature is either longer or shorter, that of the frequency feature is either more or less, that of starting hour is either earlier or later. Based on the outcome of a decision tree for anomaly detection, we extract this \textit{``@abnormal\_feature''} information. \textit{``@prediction''} describes a verb representation of an activity that has the highest transition probabilities. A sample output response is described as follows: \textit{``Alice spent much more time in going out'' and ``Alice should have slept instead of going out''} (Figure \ref{fig:system-int-diag-explain}).

The\textit{``Follow-up''} state and its child states support communication between a caregiver and an older adult (Table \ref{tab:findings-requirements}). For instance, a caregiver can request our system to confirm additional information or the status of an older adult. If a caregiver requests an inquiry (e.g. \textit{``check if he/she has a dietary problem''}) for an older adult to the system, it will store this request into the database and generate a dialogue response, such as \textit{``I will confirm whether he/she has a dietary problem''} at the \textit{``Store Request''} state. In addition, when the system recognizes the activity of an older adult that is appropriate to interact (e.g. after finishing $\textit{''SpareTime\/TV''}$ or after being $\textit{''Idle''}$), the system will prompt dialogue responses to elicit the request information from a caregiver (e.g. \textit{``I found you have an abnormal event of a toilet. I was wondering if you have any dietary problem?''}) at the \textit{``Prompt to Confirm''} state (Figure \ref{fig:system-int-diag-follow}). 

\section{Findings on Additional Use Cases \& Functionalities}\label{app_usecases}
%\subsubsection{Monitoring other fine-level daily activities}
\subsection{Monitoring other fine-level daily activities}
Both caregivers and older adults described various fine-level daily activities that are not covered in Section \ref{sect:system} but useful to be monitored by an intelligent system for older adult care. These other fine-level daily activities include medication adherence, the attempt of standing up from a bed or a chair, and time to turn around for bedridden people. 
 
%One activity that all caregivers and older adults described the most to be monitored is medication adherence. 
Caregivers and older adults described that it is important to monitor medication adherence. Older adults with chronic diseases require to take one or more medications more than once daily \cite{coleman2012dosing}. However, medication non-adherence remains a significant and widespread problem \cite{fischer2010primary} for older adults, who have cognitive demands to track medications. Older adults might \textit{``forget and past hours to take medications''} (O2) or \textit{``could not recall well whether I took my medicine or not''} (O5) and they might have \textit{``taken medicine again''} (O1). Thus, caregivers and older adults desire a system that reminds them to \textit{``not forget medication and inform times to take medication (…) especially antibiotics with schedule times''} (C1), but also \textit{``track if they (older adults) actually take medication''} (C5).
%\begin{itemize}
%    \item Attempt of standing up
%\end{itemize}
Another activity that caregivers find useful to monitor is an attempt of standing up from a bed or a chair. One leading cause of injury-related deaths for older adults is falling \cite{kannus1999fall,burns2018deaths}, which has been shown to be strongly associated with physical parameters, such as difficulty with standing from a chair \cite{rogers2003methods}. \textit{``Even though some older adults know that they can’t get up by themselves, they are more more agitated or try to get up by themselves without ringing the bell''} (C5). Thus, caregivers elaborated on the necessity of a system that can recognize the attempt of standing up by an older adult and provide an alert to a caregiver to avoid any dangerous situation (e.g. fall). 

%It would be useful to have a system to alert this event to avoid any dangerous situation (e.g. fall)''} (C5)

%\textit{``Sometimes, we forget and past hours to take medications''} (O2), and \textit{``sometimes, I could not recall well whether I took my medicine or not''} (O5). \textit{``My father used to forget that he had taken his medicine so he would take it again''} (O1).
%\textit{``Other possibilities, maybe medication takes (…) scheduled takings for certain times and alerts to not forget (…) especially antibiotics with schedule times''} (C1). \textit{``maybe it would be good to alert but you wouldn't be able to verify if they actually take it. Especially one with dementia (…) they pretend to take it (…) end up not taking it''} (C5). 

%\hfill
%
%\noindent
%\subsubsection{Environmental Monitoring: the Presence of Others \& Temperature}
\subsection{Environmental Monitoring: the Presence of Others \& Temperature}
%Older adults also described the need of monitoring environmental aspects (e.g. the presence of other people or temperature). 
Unlike most prior work that focuses on monitoring daily activities and health status of an older adult \cite{pollack2005intelligent,rashidi2012survey}, older adults provided ideas of detecting the presence of other people and environmental conditions (e.g. fire) for surveillance purposes. For instance, (O4) said \textit{``detecting whether there is a new visitor in the house of an older adult would be useful to measure the level of socialization (...) detect a dangerous invasion (e.g. thief) as surveillance''} (O4). O1 shared the experience that describes the value of monitoring environmental conditions (e.g. fire): \textit{``The medicine for my father, who had cancer, was very strong and he would be out of his mind. Once he almost set fire to his room, we got there in time, fortunately. Otherwise, he would have burned to death. (...) helpful if the system can also monitor the environmental abnormality (e.g. fire)''} (O1).

%\subsubsection{Physical assistance}
\subsection{Physical assistance}
%Caregivers and older adults shared an idea on an intelligent, robotic system that provides physical assistance on daily activities (e.g. transfer or feeding). When it comes to older adult care,
Caregivers have a great burden on providing physical assistance to older adults (e.g. transfer or feeding) \cite{adelman2014caregiver} who \textit{``are no longer independent or do have sufficient motor skills'')} (O3). This need brings opportunities for a robotic system that provides physical assistance \cite{riek2017healthcare}. For instance, caregivers and older adults desired a robotic system that can \textit{``help an older adult get up''} (O2) and \textit{``transfer an older adult from a bed to a wheelchair''} (C3) as transferring an older adult \textit{``isn’t always possible alone''} (C4). Even if there are exciting advances in robotic systems to provide physical assistance \cite{mukai2010development,erickson2018deep,gallenberger2019transfer}, it is still critical to consider and address some challenges to deploy these systems (e.g. usability, safety, and reliability issues) \cite{riek2017healthcare}. 

%\textit{``In addition to monitoring the abnormal status, it would be useful to have a system that assists people who are no longer independent or do not have great motor skills''} (O3).
%\textit{``For instance, my brother has Parkinson disease. He could not even get up by himself and need someone to help him. It would be good to have a system that can help him get up''} (O2)
    
%\subsubsection{Cognitive Assistance}
\subsection{Cognitive Assistance}
Another use case of an intelligent system is to support cognitive tasks (e.g. sending a reminder on an important event). Older adults might experience a certain type of cognitive decline: \textit{``I start forgetting most things''} (O3) or \textit{``important appointments (e.g. monthly hospital checkup''} (O5). Thus, an intelligent system that reminds to perform regular daily activities (e.g. medicine, eating, etc.) or keep medical appointments has the great potential to support the independent living of older adults \cite{pollack2003autominder,pollack2005intelligent}. In addition, as caregivers have high cognitive burdens of caring for older adults, they desire a system to reduce \red{their} cognitive burdens: \textit{``tracking and alerting the time to turn around the bedridden people (...) give foods''} (C5). 
%%%%%%%
%Pollack et al. \cite{pollack2003autominder} implemented an intelligent system that reminds about activities of daily living, conducted a preliminary field test, and found the positive excitement of older adults. However, Shishehgar et al. \cite{shishehgar2019effectiveness} described the necessity of additional studies on design and field evaluation to realize the potential of this system. 

%\subsubsection{Validations on medical symptoms}
\subsection{Validations on medical symptoms}
One of the highly appreciated potentials of our system is to provide an older adult with faster attentive care services (Section \ref{sect:finding-opp-fast}). To realize this potential better, caregivers and older adults considered that enabling a system to validate medical symptoms would be desirable. For instance, caregivers described that when older adults have abnormal, too frequent toilet events, they might just have a food problem or \textit{``urinary infections that are usually more common among older women''} (C5). Given a medical symptom, it is increasingly prevalent to search online to interpret symptoms even among older adults \cite{wald2007untangling,luger2014older}, who typically have complex needs for health care (e.g. chronic diseases or morbid conditions). \red{However,} older adults\red{, who} typically have limited health understanding or expertise \cite{berkman2011health}, could make inaccurate symptom interpretations. Thus, it would be beneficial to have an intelligent system that can \textit{``checkup the medical symptoms and status of an older adult''} (O4). 

\subsection{Communications among professional caregivers, family members, and doctors}
%\subsubsection{Communications among professional caregivers, family members, and doctors}
Along with the prior work that focuses on exploring technology designs for effective communication between care providers and adults with various chronic conditions or child patients \cite{hong2016care,berry2019supporting,seo2021learning,guan2021taking}, our study implies the necessity to explore technology designs for the communication between a professional caregiver and a family member or the communication among professional caregivers, family members, and doctors. 
%Older adults with experience in caregiving described the necessity of a system that facilitates communications among professional caregivers, family members, and doctors.

When an older adult has an emergent situation, \textit{``a family member would need to discuss with a professional caregiver and make some decisions on behalf of an older adult''} (O5).  In addition, as a professional or family caregiver might not have the medical expertise to make a prescription for an older adult, an appointment with a doctor is inevitable to provide a medical care service to an older adult. Older adults described an idea of a system that \textit{``can share the status of an older adult and receive a remote medical appointment. One time, my mother-in-law had some rashes on her body (...) waited a few hours to receive a medical diagnosis. It would be great if the system can share the symptom of an older adult by videos instead of waiting for the visiting doctor''} (O4). 

\end{document}